\documentclass{aa} % for a referee version
\usepackage{graphicx}
\usepackage{txfonts}
\usepackage{natbib}
\usepackage{caption}
\usepackage{txfonts}
\usepackage{subfigure}
\usepackage[english]{babel}
\begin{document}

\title{Searching for sub-stellar companion into the LkCa15 proto-planetary disk}
%\subtitle{ Sottotitolo }
\author{M. Bonavita \inst{1,2} \and G. Chauvin \inst{3} \and A. Boccaletti \inst{4} \and V. Pietu \inst{5} \and P. Baudoz \inst{4} \and J.~L. Beuzit \inst{3} \and A. Dutrey \inst{7} \and S. Guilloteau \inst{7} \and A.~M. Lagrange \inst{3} \and D. Mouillet \inst{3} \and G. Niccolini \inst{6}}
%1
\institute{INAF - Osservatorio Astronomico di Padova, Vicolo
  dell'Osservatorio 5, 35122 Padova, Italy 
%2
  \and Dipartimento di Astronomia, Universit\'a degli Studi di Padova, Vicolo dell'Osservatorio 2, 35122 Padova, Italy 
%3
  \and Laboratoire d'Astrophysique, Observatoire de Grenoble, UJF, CNRS, Rue de la piscine, 38400 Saint-Martin d'Hères, France 
%4
  \and LESIA, Observatoire de Paris Meudon, 5 pl. J. Janssen, 92195 Meudon, France 
%5
  \and Institut de Radio-Astronomie Millimétrique, 300 rue de la Piscine, Domaine Universitaire, 38406 Saint Martin d'Hères
%6
  \and Lab. H. Fizeau, CNRS UMR 6525, Univ. de Nice-Sophia Antipolis, Observatoire de la Côte d'Azur, 06108 Nice Cedex 2, France
%7
  \and Laboratoire d'Astrophysique de Bordeaux (LAB), Universit\'e Bordeaux 1, Bordeaux, France ; CNRS/INSU-UMR 5804, BP 89, 33270 Floirac, France }

\abstract
%context
{ Recent sub-millimetric observations at the Plateau de Bure
  interferometer evidenced a cavity at $\sim 46$ AU in radius into the
  proto-planetary disk around the T Tauri star LkCa15 (V1079 Tau), located in the
  Taurus molecular cloud. Additional Spitzer observations have
  corroborated this result possibly explained by the presence of a
  massive ($\ge 5 M_{\rm{Jup}}$) planetary mass, a brown dwarf or a low mass star companion at
  about 30 AU from the star.}
%Aims
{ We used the most recent developments of high angular resolution and
  high contrast imaging to search directly for the existence of this
  putative companion, and to bring new constraints on its physical and
  orbital properties.}
%Methods
{ The NACO adaptive optics instrument at VLT was used to observe LkCa15
  using a four quadrant phase mask coronagraph to access small angular
  separations at relatively high contrast. A reference star at the
  same parallactic angle was carefully observed to optimize the
  quasi-static speckles subtraction (limiting our sensitivity at less
  than 1.0~$\!''$).}
%Results
{ Although we do not report any positive detection of a faint
  companion that would be responsible for the observed gap in LkCa15's
  disk (25-30 AU), our detection limits start constraining its
  probable mass, semi-major axis and eccentricity. Using evolutionary
  model predictions, Monte Carlo simulations exclude the presence of
  low eccentric companions with masses $M \ge 6~M_{Jup}$ and orbiting
  at $a\ge100$~AU with significant level of confidence. For closer
  orbits, brown dwarf companions can be rejected with a detection
  probability of 90\% down to 80~AU (at 80\% down to 60~AU). Our
  detection limits do not access the star environment close enough to
  fully exclude the presence of a brown dwarf or a massive planet
  within the disk inner activity (i.e at less than 30~AU). Only, further and higher contrast
  observations should unveil the existence of this putative companion inside the LkCa15
  disk.}
%Although we do not report any positive detection of a faint
  %candidate brighter than $K_s \geq 16.5$ (and 19.5) at separation
  %respectively larger than 0.15~$\!''$ (and 0.7~$\!''$) from the star,
  %our detection limits enable us to actually constraint the probable
  %mass of a putative companion under different orbital
  %assumptions. Based on luminosity-mass relations predicted by
  %evolutionary models, we can exclude the existence of a massive
  %planet of 5, 10 and 15~$M_{\rm{Jup}}$ orbiting within the LkCa15's
  %disk at low eccentricity with probabilities of $\sim 30\%$, $95\%$
  %and $100\%$.
%Conclusions
{}


\maketitle

% 
% %%%%%%%%%%%%%%%%%%%%%%%%%%%%%%%%%%%%%%%%%%%%%%%%%%%%%%%%%%%%%%%%%%%%%%%
%%%%%%%%%%%%%%%%%%%%%%%%%%%%%%%%%%%%%%%%%%%%%%%%%%%%%%%%%%%%%%%%%%%%%%%
%%
%%

\section{introduction}
\label{sec:intro}

Since more than two decades, the first glimpse of planetary formation
has been directly observed (Smith \& Terrile 1984). Then, following
the discovery of the first extrasolar planet around a solar-type star
\citep[51 Peg, see][]{1995Natur.378..355M} , more than 400 planets have been found,
mainly using the radial velocity technique (RV). However, the
explored time span limits the radial velocity study (as well as pulsar
timing, micro-lensing, photometric transit, astrometric techniques) to
the close ($\le 4-5$ AU) circumstellar environment. To understand the
way exo-planetary systems form and evolve at wider orbits, it is then
particularly interesting to use a complementary technique, such as direct
imaging.

Around young solar analogues the current deep imaging performances enable
the detection of companions down to the brown dwarf (BD) and planetary
mass regimes.  Typical separations larger than 50-100 mas (i.e. $\ge
5-10$ AU for a star at 100 pc) can be explored for BD companions and
for planetary mass objects, typical separations in the range 200-400
mas (ie $\ge 20-40$ AU for a star at 100 pc).  The discoveries of the
planetary mass companions like 2M1207\,b (Chauvin et al. 2004) and,
more recently, around Fomalhaut \citep{2008Sci...322.1345K}, a triple
system around HR 8799 \citep{2009AAS...21331907M} and a close
candidate around $\beta$ Pic (Lagrange et al. 2009) confirm the high
contrast and high angular resolution imaging capabilities 
of current instruments. To resolve
ultra-cool and light companions, large telescopes equipped with
adaptive optics (AO) are required, as well as dedicated focal devices,
such as coronagraph and differential imagers.

Of special interest is the detection of planets still embedded in
their parental proto-planetary disk, as this may give us clues on the
timescales for planet formation and on the disk dissipation
mechanisms.  Indirect information can be given by gaps and
asymmetries observed in the disk structures as detected through
scattered light images in near infrared (NIR) \citep[e.g. for HD
  141569, see][]{1999A&A...350L..51A} or, through millimeter
interferometric measurements as recently done for AB Aur
\citep{2005A&A...443..945P} and TW Hydra
\citep{2007ApJ...664..536H}. Two controversial discoveries have been
recently obtained one with RV for the giant planet ($M \sim 10
M_{\rm{Jup}}$) at 0.04 AU in the disk that surrounds TW Hya
\citep[See][]{2008Natur.451...38S,2008A&A...489L...9H}, the other by
differential polarimetric and coronagraphic imaging for a point source
(companion or disk over-density due to dust accretion on an unseen
companion) in the outer annulus of dust of the AB Aur disk
(Oppenheimer et al. 2008). More recently, a planetary candidate has
been detected in the $\beta$ Pic circumstellar disk (Lagrange et
al. 2009).  If confirmed, these companions will directly evidence the
link between planetary formation and the main morphological and
dynamical peculiarities of transition disks.

In this paper, we present the outcomes of  NACO observations, taken
using a four quadrant phase mask coronagraph, which aims at offering
enhanced detection performances at small angular separations. Our goal
was to detect the companion responsible for the large gap, observed at mm 
wavelengths, in the disk that surrounds the young star
LkCa~15 \citep[see][]{2006A&A...460L..43P}. 

After summarizing the
properties of this star and its proto-planetary disk in section~2, we
present the NACO observations and the data reduction and analysis in
section~3 and 4 respectively. We finally report our detection limits
in section~5 and discuss them in regards of the domain of mass and
orbital parameters explored for a putative companion.

% %%%%%%%%%%%%%%%%%%%%%%%%%%%%%%%%%%%%%%%%%%%%%%%%%%%%%%%%%%%%%%%%%%%%%%%
% %%%%%%%%%%%%%%%%%%%%%%%%%%%%%%%%%%%%%%%%%%%%%%%%%%%%%%%%%%%%%%%%%%%%%%%
%%
%%

\section{LkCa 15}
\label{sec:LkCa15}

LkCa 15 (V1079 Tau: $V=12.09$,  $K=8.16$, K5 and $d \sim 140$~pc) is a T Tauri
star aged of $\sim 3-5$~Myr \citep[see][]{2000ApJ...545.1034S, 2006A&A...460..499B},
located in a hole of the Taurus molecular cloud.
\cite{2006A&A...460L..43P} led sub-mm observations of the
proto planetary disk of LkCa~15, using the Plateau de Bure
interferometer with an angular resolution of about 0.4~$\!''$. The
observations were made in ``track-sharing'', to observe LkCa~15 and
the Herbig star MWC 480 with a common calibration curve. Any
morphological difference between the two targets is thus genuine.
They found a clear inner hole around LkCa 15 while MWC 480 is
centrally peaked. This is confirmed by the modeling of the data, with
a derived inner hole of $\sim 46$ AU (i.e. the size of our own Solar
System). A multi isotopes analysis of CO rotational lines led
\cite{2007A&A...467..163P} to derive the physical properties of the
outer circumstellar disk surrounding the star. Contrary to the other
system studied in a similar way, no clear vertical temperature gradient was
found in the disk structure, possibly due to the peculiar geometry of
LkCa 15 disk.

\cite{2006A&A...460L..43P} discussed the possible mechanisms that
could explain the inner hole structure \citep[not completely empty, as
  some IR excess is present above the stellar black
  body. See][]{2004ApJ...614L.133B}.  Planetary formation or the
presence of a low-mass companion seems to be plausible since the
LkCa15 disk is quite massive.  In fact \cite{2000ApJ...545.1034S}
estimated, from the kinematics, a total mass for the system of
$M_{\rm{dyn}}=1.00\pm 0.1 M_{\odot}$. This set the maximal mass of
the putative companion to $0.2~M_{\odot}$, since the mass of LkCa 15
could hardly be less than $0.8~M_{\odot}$, due to its spectral type.
This value is in agreeement with a recent estimate of the mass of the star
obtained by \cite{2007A&A...473L..21B}. 
Using a new distance estimate by \cite{2006A&A...460..499B},
they confirmed LkCa 15 as a kinematic member of the Taurus association 
seen through a hole in the molecular cloud, retrieving a mass value 
of $1.12\pm 0.08 M_{\odot}$.

Additional results obtained by \cite{2006MNRAS.369..229A} also suggest
that other mechanisms such as photo-evaporation are unlikely to be
effective due to the large disk density, which is significantly higher
than the one expected if the photo-evaporation starts propagating
beyond 20-30 AU. \cite{2006A&A...460L..43P} suggest that a $\sim 5-10
M_{\rm{Jup}}$ planet orbiting at 30~AU would be sufficient to evacuate
the inner 50 AU of the LkCa 15 disk. Similar conclusions were reached
 by \cite{2007ApJ...670L.135E} with Spitzer
observations and the modeling of the SED of LkCa 15. Their analysis
suggests that a gap is present in the disk of LkCa 15, with an inner
disk going from 0.12-0.15 AU to 4-5 AU, and an outer disk of inner
radius 46 AU, in perfect agreement with the findings of
\cite{2006A&A...460L..43P}. They also concluded that planetary
formation or the presence of a close-in stellar or sub-stellar
companion are the most probable explanations for the circumstellar
material shape around LkCa15 

%\begin{table}[t]
%\centering
%\caption{\footnotesize Physical parameters for the LkCa15
%  proto-planetary disk reported by \cite{2007A&A...467..163P}}
%\begin{tabular}{lc}
%\noalign{\smallskip}\noalign{\smallskip}
%\hline
%\hline
%\noalign{\smallskip}\noalign{\smallskip}
%Orientation, PA (deg)                   & $151 \pm 3$ \\ 
%Inclination, i  (deg)                   & $49 \pm 3$ \\
%$R_{Int}$ (AU)                              & $46 \pm 3$ \\
%$R_{Out}$ (AU)                              & $177 \pm 12$ \\
%Surface density at 100 AU (g.cm$^{-2}$)   & $3.1 \pm 0.4$ \\
%$M_{Disk}$  $(M_{\odot})$                  & 0.029 \\
%Temperature at 100~AU (K)               & $22\pm 1$ \\
%\noalign{\smallskip}\noalign{\smallskip}
%\hline
%\hline
%\end{tabular}
%\label{tab:diskprop}
%\end{table}

% %%%%%%%%%%%%%%%%%%%%%%%%%%%%%%%%%%%%%%%%%%%%%%%%%%%%%%%%%%%%%%%%%%%%%%%
% %%%%%%%%%%%%%%%%%%%%%%%%%%%%%%%%%%%%%%%%%%%%%%%%%%%%%%%%%%%%%%%%%%%%%%%
%%
%%

\section{Observations}
\label{sec:Observations}

% %%%%%%%%%%%%%%%%%%%%%%%%%%%%%%%%%%%%%%%%%%%%%%%%%%%%%%%%%%%%%%%%%%%%%%%
%%
%%

\subsection{Telescope and instrument}
\label{sec:telescope}

The observations were performed on December 26th, 2007 at ESO/Paranal,
using NACO, the AO-assisted near-IR camera NAOS-CONICA
\citep{2003SPIE.4839..140R} mounted on one of the Nasmyth focus of the
UT4 8m-telescope. Among the numerous NACO observing modes
\citep{2003SPIE.4841..944L}, the classical and coronagraphic imaging
modes were used.

Coronagraphic observations were performed with the four-quadrant phase
mask (4QPM) optimized for $K_s$ band observations. The S13 objective
(FoV of $14~\!'' \times 14~\!''$ and plate-scale of 13.25 mas/pixel)
was chosen for a more precise centering with the 4QPM and a better
sampling of the PSF. The 4QPM splits the focal plane into four equal
areas, two of which are phase-shifted by $\pi$.  As a
consequence, a destructive interference occurs in the relayed pupil
and the on-axis starlight rejected on the edge of the geometric pupil
is filtered with a Lyot stop, being a circular hole 90\% 
of the pupil size.
The advantage over the classical Lyot mask is the possibility to
access inner angular separations lower than $0.35~\!''$ (the smallest
NACO occulting mask) at relatively large contrast
\citep[see][2008]{2004PASP..116.1061B}.

However, a significant part of the starlight is left in the focal
plane due to uncorrected aberrations composed of a dynamical halo
averaging over time plus a quasi-static halo corresponding to optical
aberrations along the optical train (from telescope to detector).  To
mitigate this problem, a coronagraphic image of a reference star was
taken just after our science target observations, with the same
instrumental settings, to serve for speckle calibration in an
image-subtraction process.  This reference star (BD + 22 729, $V
=11.5$, $K= 7.9$) has similar visible and NIR magnitudes to ensure
similar AO correction and signal-to-noise at the detector. Moreover,
observations were performed to match the parallactic angle with
LkCa15, i.e the instrumental pupil configuration, in order to optimize
the overlap of the speckle pattern and diffraction spikes position in
the final subtracted image of LkCa15.

\begin{table}[t]
%\centering
\caption{\label{t7} Observing parameters used for each source,
  observed in classical and coronagraphic imaging.}
\begin{tabular}{lrrr}
\hline
\hline
\noalign{\smallskip}
Object                                                  &  LkCa15        & LkCa 15         & BD +22 729              \\
                                                        &  (Classical)   & (Coronagr.)  & (Coronagr.)          \\
\noalign{\smallskip}
\hline
\noalign{\smallskip}
Filter                                                  & $ND+NB_{2.17}$ & $K_s$           & $K_s$                   \\
Objective                                               &  S13           & S13             & S13                     \\
$t_{\rm{int}}$  (s)                    &  30            & 24              & 24                      \\
$N_{frame}\times N_{exp}$              &  $2\times1$    & $4\times8$      & $4\times8$                     \\
$t_{\rm{int}}[\rm{sky}]$  (s)          &  30            & 24              & 24                      \\
$N_{frame} \times N_{exp}[\rm{sky}]$   &  $2\times1$    & $4\times1$      & $4\times1$                     \\
\noalign{\smallskip}
\hline
\hline 
\end{tabular}
\tablefoot{ \bf The individual integration time ($t_{\rm{int}}$), 
  the number of frames ($N_{frame}$) 
  averaged by the detector and the number of repeated exposures 
  ($N_{\rm{exp}}$) are reported on source and on sky. 
  At the end, $N_{frame}\times N_{\rm{exp}}$ gives the
  number of images available for the data reduction and analysis. }

\label{tab:obs1}
\end{table}

% %%%%%%%%%%%%%%%%%%%%%%%%%%%%%%%%%%%%%%%%%%%%%%%%%%%%%%%%%%%%%%%%%%%%%%%
%%
%%

\subsection{Observing strategy}
\label{sec:obs_strat}

The coronagraphic observations were preceded by a classical imaging
sequence that provides a photometric reference. A neutral density
(1.12\% transmission) and a narrow band filters at 2.17~$\mu$m were
used to serve as an image quality check and as photometric reference
for the coronagraphic observations. Table~\ref{tab:obs1} summarizes
the observing parameters.  During the coronagraphic observing
sequence, the precise centering of the science target behind the focal
plane mask was critical to maximize the central star attenuation. The
coronagraphic observations were acquired at two instrument positions
rotated by 33$^{\circ}$. Sky images were immediately observed for
LkCa15 and its reference, after each coronagraphic sequence.

% %%%%%%%%%%%%%%%%%%%%%%%%%%%%%%%%%%%%%%%%%%%%%%%%%%%%%%%%%%%%%%%%%%%%%%%
% %%%%%%%%%%%%%%%%%%%%%%%%%%%%%%%%%%%%%%%%%%%%%%%%%%%%%%%%%%%%%%%%%%%%%%%
%%
%%

\section{Data reduction and analysis}
\label{sec:data_red}

% %%%%%%%%%%%%%%%%%%%%%%%%%%%%%%%%%%%%%%%%%%%%%%%%%%%%%%%%%%%%%%%%%%%%%%%
%%
%%

\subsection{Image processing and selection}
\label{sec:cosmetic}

The data were processed using the
Eclipse \footnote{http://www.eso.org/projects/aot/eclipse/} reduction
software \citep{1997Msngr..87...19D} for bad pixel correction, flat
fielding and sky subtraction. Individual frames were inspected by eye
to remove all low-quality images degraded by waffle aberrations or
variable AO corrections. Finally, only LkCa15 and BD + 22 729 images
with similar parallactic angles were selected to optimize our
PSF-subtraction. Among the initial data set of 8 images per source and
per rotator position, only 3 were kept after the selection process
totaling an integration time of 288 seconds.

% %%%%%%%%%%%%%%%%%%%%%%%%%%%%%%%%%%%%%%%%%%%%%%%%%%%%%%%%%%%%%%%%%%%%%%%
%%
%%
\subsection{Subtraction of the diffraction residuals}
\label{sec:psfsub}

\begin{figure}[tbp]
\centering
\includegraphics[width=\columnwidth]{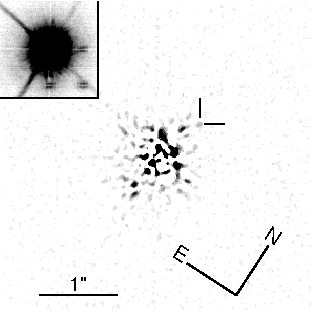}\qquad\qquad\qquad\qquad
\caption{LkCa15 4QPM coronagraphic image at a rotator
    position of 33$^{\circ}$ after PSF
    subtraction and spatial filtering. The zoom-in non-subtracted image is reported on the upper-left corner with the same field of view and thresholds.}
\label{fig:sub0}
\end{figure}

  The purpose is to properly subtract the stellar contribution from
  the LkCa15 images. For each rotator and parallactic angle position,
  the reference star was shifted at a 0.1-pixel accuracy, scaled and
  subtracted to minimize the residuals. An IDL custom-made tool was
  used to rapidly converge on an acceptable shift and scaling
  solution. Alternatively, a residual minimization using the
  AMOEBA\footnote{\tiny
    $http://www.physics.nyu.edu/grierlab/idl\_html\_help/A8.htmlwp992475$}
  function was applied giving consistent results. 
  Fig.~\ref{fig:sub0} shows the result of the single image-subtraction
  image of LkCa15 at a rotator position of 33$^{\circ}$ corresponding
  to the set that provides the best contrast. Additionally, we took
  advantage of the observations taken at two rotator positions to
  explore dead zones hidden by the secondary spikes and the
  coronagraphic mask transition (see zoom-i, in Fig.~1). Derotation was also applied to
  average the speckles patterns and increase the final
  signal-to-noise. Finally, Double subtraction of images at 0 and
  33$^{\circ}$ were obtained to remove the non rotating aberrations
  (related to static optics in NACO) and possibly reveal a
  positive-negative signature expected for a true companion (and not
  for instrumental residuals).  All sets of single, double-subtracted
  and derotated-averaged images were finally compared to derive a
  final subtracted image, optimized at different angular separations.

% %%%%%%%%%%%%%%%%%%%%%%%%%%%%%%%%%%%%%%%%%%%%%%%%%%%%%%%%%%%%%%%%%%%%%%%
%%
%%

\subsection{Detection limit}
\label{sec:detlim}

Using our final residual image, a pixel-to-pixel 2D noise map was
estimated using a sliding box of $5\times5$ pixels over the whole NACO
FoV (see Fig.~\ref{fig:2D_deltaK} and 1D-detection limits in  Fig.~\ref{fig:1D_deltaK}).  The 6~$\sigma$ detection limit map was obtained
after renormalization by the LkCa15 images obtained in classical
imaging and applying all corrections related to the use of different
optical set-ups: exposure times (see Table~\ref{tab:obs1}), neutral
density and Lyot-stop transmissions \citep[a factor of 1.12\% and
  80.8\% respectively, according to][]{2008A&A...482..939B} and
$NB_{2.17}$ to $K_s$ filter transformation. Finally, we also took into
account that the presence of the four-quadrant mask causes an
attenuation of the off-axis objects due to the 4QPM transition,
decreasing our sensitivity close to the axis
\citep[see][]{2004PASP..116.1061B}. Compared to previous deep imaging
surveys with no a-priori information on the orbital configuration of
the system, the LkCa~15 disk system properties offer the possibility
to set constraints on the orbital properties of a possible co-planar
companion. The use of the complete 2D-detection map is therefore
mandatory to fully exploit the sky-projected spatial information of the LkCa 15 environment provided by our direct imaging observations.

% %%%%%%%%%%%%%%%%%%%%%%%%%%%%%%%%%%%%%%%%%%%%%%%%%%%%%%%%%%%%%%%%%%%%%%%
% %%%%%%%%%%%%%%%%%%%%%%%%%%%%%%%%%%%%%%%%%%%%%%%%%%%%%%%%%%%%%%%%%%%%%%%
%%
%%

\section{Results}
\label{sec:results}

% %%%%%%%%%%%%%%%%%%%%%%%%%%%%%%%%%%%%%%%%%%%%%%%%%%%%%%%%%%%%%%%%%%%%%%%
%%
%%

\subsection{Null-detection in the central hole}
\label{sec:null_res}

\begin{figure}[tbp]
\centering
\includegraphics[width=\columnwidth]{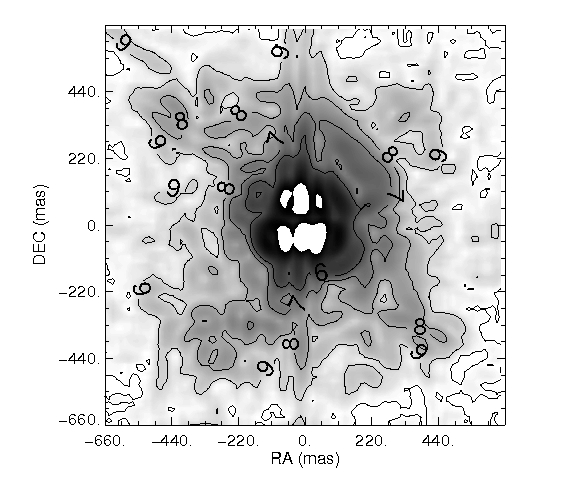}
\caption{$\Delta K_{S}$ contrast map giving
    the contrast between the faintest target detectable at a $6\sigma$
    level and the primary star. }

\label{fig:2D_deltaK}
\end{figure}

\begin{figure}[tbp]
\centering
\includegraphics[width=\columnwidth]{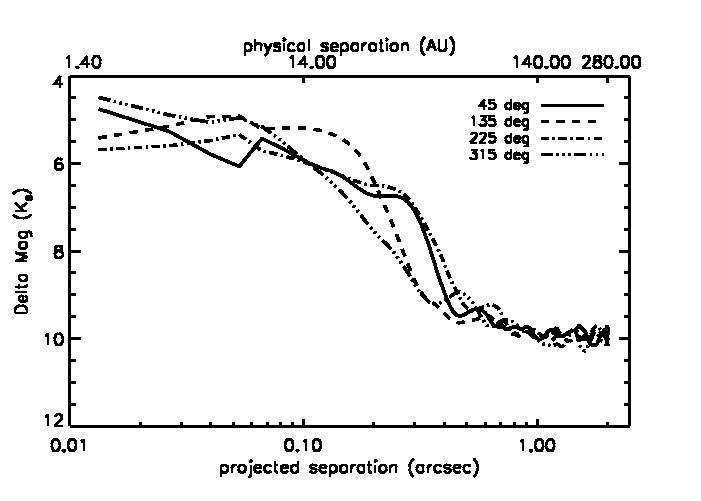}

\caption{As a comparison are reported
    four different $\Delta K_{S}$ 1D-contrast detection limits
    extracted at 4 position angles (45, 135, 225 and 315
    degrees). They illustrate the azimuthal variation at close inner
    angles due to the NACO PSF structure and the relevancy of using a
    contrast map to take that effect into account.}

\label{fig:1D_deltaK}
\end{figure}

At separations corresponding to the central hole ($\rho<0.33~\!''$, 46~AU)
no evidence of a point-like object was found in any of our
image-subtracted sets. The azimuthal averaged contrast places a limit
of $\Delta K_s=7$mag equivalent to about $12 M_J$ for a hypothetical
projected distance of 30\,AU. A statistical approach of this detection
limit is analyzed thoroughly in section \ref{sec:companion}. We also
searched for point-like objects at separations that are not compatible
with the presence of the central hole. We found a low significant
point-source at a separation of $\rho= (0.67\pm 0.02)"$ at a position
angle of $PA=(340.7\pm 0.3)^{\circ}$. Although close to the detection
limit, this point-source lies at the boundary between the speckle and
the background noise regimes and is then visible in the subtracted
image of Fig.~\ref{fig:sub0}.  We coarsely estimated a contrast of
$\Delta K_s=10.2$mag in a 5 pixels aperture. Error bars are not
estimated as the point-source is quite close to the noise level (less
than 6$\sigma$ if we refer to Fig.~\ref{fig:1D_deltaK}), so the
previous contrast value should be taken with caution. There are a
number of artifacts that may produce such patterns like waffle and
spiders. The nearest spider spike of which the trace is still visible
in Fig.~\ref{fig:sub0} is offset by $25^{\circ}$ while the waffle mode
appears in a $45^{\circ}$ direction at a position of $0.49~\!''$ therefore
not compatible with the presence of this point-source. However, it is
not visible in the subtracted image obtained for the 0$^{\circ}$
rotator position while at such separation the detection limit is
almost similar for both rotator positions (see
Fig.~\ref{fig:2D_deltaK}). But the point-source would be at about
$0.2~\!''$ from the vertical 4QPM transition and slightly attenuated. As it
is difficult to rule out a structure in the speckle pattern,
additional observations will be mandatory to test the presence of this
potentially low-mass object.

% %%%%%%%%%%%%%%%%%%%%%%%%%%%%%%%%%%%%%%%%%%%%%%%%%%%%%%%%%%%%%%%%%%%%%%%
%%
%%

\subsection{Companion mass and orbital parameters}
\label{sec:companion}

\subsubsection{Simulation description}
\label{sec:sim_description}

A first step is to convert our 6~$\sigma$ detection limit map in terms
of minimum mass map. This is usually done considering the star
apparent magnitude in $K_s$-band and its distance, finally the
evolutionary model predictions at the age of the system to convert the
absolute magnitude limits derived in the NACO filters in masses.  We
have considered here an age of 4~Myr and a distance of 140~pc for
LkCa15. We then considered two classes of evolutionary models based on
different assumptions on the initial conditions: Hot Start models
considering an initial spherical contracting state
\citep{2000ApJ...542..464C, 2003A&A...402..701B, 2008ApJ...689.1327S}
and core accretion models coupling planetary thermal evolution to the
predicted core mass and thermal structure of a core-accretion planet
formation model \citep{2007ApJ...655..541M, 2008ApJ...683.1104F}.  In
the case of the core-accretion model predictions, our detection
performances do not allow to access the planetary mass regime at
all. Massive hot Jupiters are indeed predicted to be much fainter at
young ages \citep{2007ApJ...655..541M}. Therefore, minimum mass maps
were determined using Hot Start model predictions over the planetary
and brown dwarf mass regime.
  
In a second step we used our minimum mass maps to calculate the
detection probability ($P_{\rm{D}}$) of companions of various masses
and orbital parameters (semi-major axis $a$, eccentricities $e$,
inclination $i$, longitude of the ascending node $\Omega$, longitude
of periastron $\omega$ and time of periastron passage
$T_p$). Considering the disk properties (inclination, position angle
and inner radius), different assumptions can be made to fix partially
the companion orbital properties. The rest of the orbital parameters
can be randomly generated.  Each simulated companion is then,
according to its position on the projected orbit, placed on our 2D
minimum mass map, and its detectability is tested, comparing its mass
with the minimum value achievable at the same position in the FoV.
Running the simulation $10.000$ times for a given set of mass and
orbital parameters enables to derive a detection probability.  As
already pointed out in Sec. \ref{sec:detlim}, we took advantage of
both the available information on the disk geometry plus the spatial
information coming from the detection limit map, in order to test all
probable set of physical and orbital parameters for a putative
companion.
  
The smallest projected physical separation probed around LkCa15 is
limited by the 4QPM coronagraph attenuation inside 0.15$~\!''$ (equivalent
to 3 times the angular resolution), setting the minimum semi-major
axis considered in our simulations to 20 AU.  Projected physical
separations as large as 1000~AU are explored, but we decided to
restrain our study to the close circumstellar environment considering
semi-major axis $a = [20,280]$~AU. Additionally, we restrain the
parameters space explored for $M=[3,100]~M_{\rm{Jup}}$ and
eccentricity $e=[0.,0.3]$.
  
\begin{figure*}[t]
\centering
\hspace{-1.cm}
\includegraphics[width=11cm]{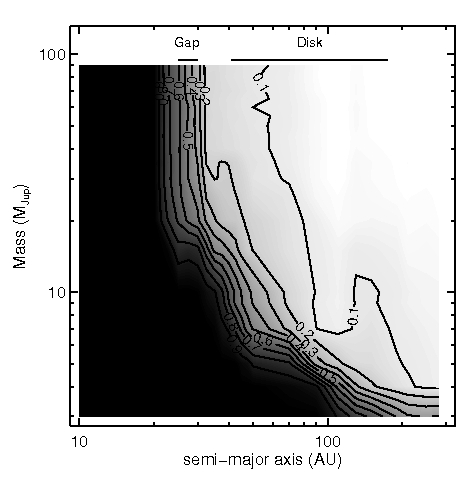}\qquad
%\captionsetup{raggedright}

\caption{Non-detection probability map of a faint companion as a
  function of its mass and semi-major axis in the case of a
  null-eccentricity (circular) orbit solution. Inclination and
  longitude of the ascending node have been fixed using the disk
  properties: $i_{Disk}=49^o$ and $\Omega = PA_{Disk} \pm 90^o \left(=
  151\pm90\right)$.}
\label{fig:simu}
\end{figure*}

\subsubsection{Results}
In Fig.~\ref{fig:simu}, the results of our simulation are given in
terms of non-detection probability map as a function of the companion
mass and semi-major axis. We considered the most constraining case of
a coplanar orbit with an inclination and a longitude of the ascending node
fixed by the disk properties: $i_{Disk}=49^o$ and $\Omega = PA_{Disk}
\pm 90^o \left(= 151\pm90^o\right)$. Only the null-eccentricity map is
showed, but the solutions for low-eccentricities ($e<0.3$) are not
significantly different. The range of semi-major axis estimated for
the predicted candidate companion \citep[25-35 AU,
  see][]{2006A&A...460L..43P} and the disk boundaries ($46-177$ AU) are reported,
as reported by \cite{2007A&A...467..163P}.
  
Any low stellar mass companion with masses should have been detected
with a high confidence level (non-detection probability lower than
0.1) for semi-major axis larger than 55~AU (non-detection probability
of 0.2 at 30~AU). Any brown dwarf companion with masses larger than
$10~M_{Jup}$ should have been detected for semi-major axis larger than
80~AU with a non-detection probability of 0.1 (60~AU for 0.2). Our
non-detection probability map shows that we start exploring the
planetary mass regime but not deep enough to reject the presence of
any massive giant planet at the location of the disk depletion (non
detection probability higher than 0.9 at 20--35~AU). For this physical
separation, the presence of more massive companions, 20 $M_{Jup}$ or
above, cannot be significantly excluded.  At larger distances
($a>100~AU$), the presences of companions of any mass above
$6~M_{Jup}$ seem unlikely. Although we only explored low eccentricity
solutions in a systematic way, the above results also provide lower
limits on the detection probabilities for larger eccentricity values,
as the time spent far from the star becomes progressively larger.

\subsubsection{Limitations}

As for all the current deep (or even wide field) imaging studies of
young, very low mass objects or companions \citep[See
  e.g.][]{2009arXiv0909.4531N,2009arXiv0906.2945C,2007ApJ...670.1367L,2008Sci...322.1348M},
the main limitations of our analysis come from the conversion of our
detection limits in terms of minimum detectable masses. They are
related to the uncertainty in the age determination of LkCa15 and the
use of non-calibrated evolutionary models for young ages and very low
masses. By comparison, uncertainties on the system distance and $K_s$
apparent brightness are negligible. Age and model predictions are
discussed below:

\begin{enumerate}

\item LkCa15 is a confirmed member of the Taurus-Auriga association,
  for which an age of 3--5~Myr is estimated \citep{2000ApJ...545.1034S}. 
  To explore how the age uncertainty affects our results, 
  we ran similar simulations for ages of 1, 4 and 7~Myr.  
  The impact of the age uncertainty is much more important in the 
  planetary mass domain than in the brown dwarf regime. 
  Consequently, it does not significantly affect our conclusions 
  relative to the presence of a brown dwarf or low stellar mass 
  companion around LkCa15.

\item The applicability of evolution tracks of
  brown dwarfs at ages less than a few million years have been already
  cautioned by \cite{2003A&A...402..701B}. The role of the initial
  conditions has been also questioned by \cite{2007ApJ...655..541M},
  specifically for young giant planets where a connection between
  giant planet formation and evolution models seems mandatory. At the
  location of the LkCa15 inner disk cavity (between 20 and 46~AU), our
  4QPM observations are not sensitive at all to apparent magnitudes
  predicted for planetary mass companions described by the core
  accretion start models. They are marginally sensitive for high
  planetary masses with fluxes predicted by the Hot Start models. The
  strongest constraints are actually set in the brown dwarf regime for
  masses larger than $M\ge20~M_{Jup}$, where the Hot Start evolutionary models
  are more appropriate. Although these models still need to be more
  extensively calibrated for this range of masses and ages, recent
  discoveries of young calibrators \citep{2004ApJ...609L..33M, 2005Natur.433..286C, 2006noao.prop...99S} indicate that their predictions are
  relatively faithful regarding the accuracy required in our
  study. Our conclusions related to the probable presence of a brown
  dwarf or low mass star companion responsible for the inner hole
  detected in the LkCa15 disk remain then meaningful down to 30~AU.

\end{enumerate}

\section{Conclusions}
\label{sec:conclusions}

The T-Tauri star LkCa 15 was observed with VLT/NACO using the 4QPM
coronagraph, reaching a contrast lower than 9.5 in $K_s$ band, at
separations higher than 0.5~$\!''$.  Our goal was the detection of a
low-mass companion, with a mass spanning from 0.2~$M_{\odot}$ down to
5~$M_{Jup}$, which presence has been suggested as an explanation for
the large cavity, evidenced by sub-millimeter observations, in the
disk surrounding the star. 

We do not report any positive detection of
a close companion to the star LkCa15. Based on our detection limits
and Hot Start evolutionary model predictions, we ran simulations to
take into account that the presence of a putative companion enable us
to constrain its mass and semi-major axis given reasonable assumptions
from the disk geometry.

  The innovative approach in our case is the combined use of either
  the available information on the disk configuration (inclination,
  ascending node, gap prediction) and the complete 2D information of
  the coronagraphic image, to actually constrain the probability of
  existence of a stellar or sub-stellar companion around LkCa15.
  As a main result of our statistical analysis we can exclude the
existence of a low mass star and brown dwarf companion  with semi-major
axis larger than 55 and 80~AU respectively. The planetary mass regime
is only partially constrained and at large semi-major axis
($a \ge 100$~AU) the existence of massive ($M \ge 6~M_{Jup}$) planetary
mass companions seem unlikely.

The limitations of our study is also discussed in regards of the
LkCa15 age uncertainty that does not affect much our conclusions in
the case of low mass star and brown dwarf companions, and of the
uncertainty related to the initial conditions adopted for the
evolutionary model, particularly critical in the planetary mass
regime. Finally, in case of a true companion orbiting LkCa15 in the
inner disk cavity, our observations would favor a planetary mass or
low-mass brown dwarf companion, although more massive companions
cannot be completely excluded with high detection probability.

Further deep imaging studies at the 5--50~AU scale at a new epoch
should provide complementary information to completely reject the
existence of close stellar or brown dwarf companion to LkCa15 and
pursue the search for the putative companion that would be responsible
for the disk geometry and inner cavity within 46~AU.

\bibliographystyle{aa}
\bibliography{lkca15_revised_II}

\end{document}